\documentclass[aps,10pt,a4paper,amsfonts,nofootinbib,onecolumn]{revtex4}
\usepackage{graphicx}
\usepackage{mathrsfs}
\newcommand{\gapp}{\mathrel{\vcenter{\hbox{\tiny\ooalign
{\raise 3.25pt\hbox{$>$}\crcr $\sim$}}}}}

\renewcommand{\mathbf}[1]{{\mbox{\boldmath $\mathrm{#1}$}}}%{{\mathrm{\bf #1}}}

\newcommand{\be}{\begin{equation}}
\newcommand{\ee}{\end{equation}}
\newcommand{\ba}{\begin{eqnarray}}
\newcommand{\ea}{\end{eqnarray}}

\newcommand{\bac}{\begin{equation}
    \begin{array}{rcl}}
\newcommand{\eac}{\end{array}\end{equation}}

\newcommand{\bra}[1]{\left(#1\right)}
\newcommand{\bras}[1]{\left[#1\right]}
\newcommand{\brac}[1]{\left\{#1\right\}}

\renewcommand{\:}[2]{{\textstyle\frac{#1}{#2}}}

\newcommand{\forget}[1]{\iffalse#1\fi}
\newcommand{\forgetmenot}[1]{\iftrue#1\fi}

\newcommand{\del}{\nabla}

\renewcommand{\div}{\hskip0.9pt{\mathsf{div}\hskip2pt}}
\newcommand{\curl}{\hskip0.9pt{\mathsf{curl}\hskip2pt}}
\newcommand{\dis}{\hskip0.9pt{\mathsf{dis}\hskip2pt}}
\newcommand{\sdel}{{\mathrm{D}}}

\newcommand{\<}{\langle}
\renewcommand{\>}{\rangle}

\renewcommand{\S}{_{\hskip-0.8pt\mathrel{\vcenter{\hbox{\tiny\ooalign
{\raise 1.5pt\hbox{\textsf{S}}}}}}}}
\newcommand{\V}{_{\hskip-0.8pt\mathrel{\vcenter{\hbox{\tiny\ooalign
{\raise 1.5pt\hbox{\textsf{V}}}}}}}}
\newcommand{\T}{_{\hskip-0.8pt\mathrel{\vcenter{\hbox{\tiny\ooalign
{\raise 1.5pt\hbox{\textsf{T}}}}}}}}

\renewcommand{\b}[1]{{\mbox{\boldmath $\mathrm{#1}$}}}

\begin{document}

\title{{\textrm{\Large\MakeLowercase{
\sc Cosmological density fluctuations and gravity waves: A
covariant approach to gauge-invariant non-linear
perturbation~theory }}}}

\author{\vspace{2mm} Chris~Clarkson}

\affiliation{\vspace{2mm}Relativity and Cosmology Group,
Department of Mathematics and Applied Mathematics, University of
Cape Town, Rondebosch 7701, Cape Town, South Africa;\\ Institute
of Cosmology and Gravitation, University of Portsmouth,
Portsmouth, PO1 2EG, Britain
\vspace{2mm}}
\email{chris.clarkson@port.ac.uk}

\date{\small{\today}}

%\maketitle

\begin{abstract}

We present a new approach to gauge-invariant cosmological
perturbations at second order, which is also covariant.  We
examine two cases in particular for a dust Friedman-Lema\^\i
tre-Robertson-Walker model of any curvature: we investigate
gravity waves generated from clustering matter, that is, induced
tensor modes from scalar modes; and we discuss the generation of
density fluctuations induced by gravity waves~-- scalar modes from
tensor perturbations.  We derive a linear system of evolution
equations for second-order gauge-invariant variables which
characterise fully the induced modes of interest, with a source
formed from variables quadratic in first-order quantities; these
we transform into fully-fledged second-order gauge-invariant
variables. Both the invariantly defined variables and the key
evolution equations are considerably simpler than similar
gauge-invariant results derived by other methods. By finding
analytical solutions, we demonstrate that non-linear effects can
significantly amplify or dampen modes present in standard
linearised cosmological perturbation theory, thereby providing an
important source of potential error in, and refinement of, the
standard model. Moreover, these effects can dominate at late
times, and on super-Hubble scales.
%In particular, such
%damping of the spectrum of tensor modes by scalar modes may
%provide a part of the explanation for the unexpected absence of
%gravitational waves imprinted in the cosmic microwave background
%ecently reported by the WMAP team.

%\pacs{}
\end{abstract}

\maketitle

\section{\textmd{{{ introduction}}}}
%\sec{introduction}

The standard model of cosmology is built using perturbations of
the homogeneous and isotropic Friedman-Lema\^\i
tre-Robertson-Walker (FLRW) models. These perturbations have, with
a few exceptions, been limited to first-order thus far; at
first-order, the generic large-scale features of the universe can
be accounted for. However, there are good reasons for
investigating higher-order perturbations as a refinement of the
accuracy of this model. The unprecedented precision of present and
upcoming cosmic microwave background (CMB) measurements, such as
WMAP and Planck, are reaching levels where non-linear effects may
play an important role, especially in any non-Gaussianity present.
Indeed, it is rapidly becoming an important area of research to
try to predict possible imprints of quantum gravity effects in the
CMB; but this can only properly succeed if we also fully
understand the range of classical non-linear effects which may
produce very similar fluctuations. Non-linear effects may also
play an important role in inflation as in the stochastic approach
to inflation, which partly includes second-order back-reaction
effects~\cite{GLMM}, and also from a fully second-order
approach~\cite{ABMR2}.

Another reason to consider second-order perturbations is that
there is no way, within linear theory, to decide when the
perturbations have become too large for the theory to handle~--
i.e., when non-linearities should be taken into account. The
reason is that, while it would seem sensible to compare the
magnitudes of linearised objects to some invariant background
variable, such as the energy density, say, this does not always
work, a simple example being gravitational waves in flat space.
%: for example, at late times in an open or flat model the
%energy density tends to zero at late times, but linear
%perturbation theory is still a perfectly acceptable description of
%gravity waves whose amplitude may then be much larger than the
%energy density.
The study of cosmology is not yet at an advanced enough stage for
fully non-linear numerical predictions, as is the case for many
astrophysical situations, so second-order perturbation theory is
the tool which must be developed for this.

For these reasons, second-order cosmological perturbation theory
is beginning to be investigated, following work in the sixties by
Tomita~\cite{tomita} who first investigated the second-order terms
from scalar modes. For example, other authors have considered
second-order perturbations of a flat dust FLRW models, both
with~\cite{MTB} and without~\cite{MMB} a cosmological constant.
Inflation has also been considered at second order~\cite{ABMR2},
providing the prediction of the bispectrum of perturbations from
inflation. Meanwhile, the evolution of the curvature perturbation
on super-Hubble scales after inflation has been shown to be
constant at second order~\cite{MW,BCLM}. More recently, it has
been shown that second-order effects can lead to detectable
non-Gaussianity in the CMB~\cite{BMR1,BMR2}, as well as
polarization~\cite{MHM}. In this paper we complement these results
by demonstrating that second-order effects can dominate at late
times and on super-Hubble scales over first-order modes.

A crucial aspect of relativistic perturbation theory is the
mapping, or gauge choice, between the (fictitious) background and
the perturbed spacetime; many perturbation approaches are not
invariant with respect to this mapping (see~\cite{SW,BS} for a
comprehensive discussion of this). In terms of coordinates,
objects are typically not invariant under infinitesimal coordinate
changes. This means that objects and modes specified in one gauge
do not always correspond to objects and modes in another, making
the interpretation of variables, and the physical situation,
difficult. In fact, a gauge-dependent object is not
observable~\cite{BS}, which follows from the principle of general
covariance. It is therefore important to have a method of
describing perturbations using variables which are invariant under
gauge transformations. A gauge-invariant object at
\emph{first-order} is a quantity which vanishes (or is constant)
in the background; this is the Stewart-Walker Lemma~\cite{SW}.
This gauge problem was first solved in cosmology by
Bardeen~\cite{bardeen} by providing a full set of gauge-invariant
quantities with which to describe the perturbed spacetime.
Bardeen's variables are formed from gauge-invariant (GI) linear
combinations of gauge-dependent variables, given in a particular
coordinate system; but this prescription can make their
interpretation difficult.

This situation has been clarified more recently by the covariant
and gauge-invariant perturbation approach of Ellis, Bruni, and
others~\cite{EB,BDE}, using the covariant 1+3 approach~\cite{EvE}.
Its strength in cosmological applications lies in the fact that it
is well adapted to the system it is describing: all essential
information can be captured in a set of (1+3) covariant variables
(defined with respect to a preferred timelike observer congruence
$u^a$), that have an immediate physical and geometrical
significance. These variables satisfy a set of {evolution} and
{constraint} equations, derived from Einstein's field equations,
and the Bianchi and Ricci identities, which form a closed system
of equations when an equation of state for the matter is chosen.
The covariant and gauge-invariant linearisation procedure is easy
and transparent: it consists of deciding which variables are
`first order' (or `of order $\epsilon$') and those which are
`zeroth order'~-- i.e., those which do not vanish in the
background, which is usually a FLRW model. Products of first-order
quantities can then be ignored in the equations.
%All
%projected vectors and tensors are first-order, so there is no
%vector-tensor and tensor-tensor coupling in the equations.
%Harmonic functions can then be introduced which re-write the
%equations in scalar form; the resulting system is then in the form
%of algebraic constraints and some {first-order} ordinary
%differential equations; the solution is then straightforward.
The key point of the approach is that it deals with physically or
geometrically relevant quantities, such as the fractional density
gradient, ${\cal D}_a=a\sdel_a\mu/\mu$ (where $a$ is the scale
factor, $\mu$ is the gauge-dependent energy density in the
perturbed spacetime, and $\sdel_a$ is the derivative operator in
the observers' rest space), and the comoving spatial expansion
gradient ${\cal Z}_a=a\sdel_a\theta$ (where $\theta$ is the
gauge-dependent expansion), being the most important for scalar
perturbations, while the electric and magnetic parts of the Weyl
tensor, $E_{ab}$ and $H_{ab}$, respectively, represent the
non-local parts of the gravitational field, and describe, amongst
other things, the propagation of gravitational waves.  All
3-vectors and projected, symmetric, trace-free tensors in this
approach, are automatically GI because of the symmetry of the
background.

The gauge problem at second-order is quite a challenge: gauge
transformations at second-order for second-order variables are
\emph{huge} in general~-- often as complicated as the evolution
equations themselves~(see e.g., \cite{MMB,BS,ABMR2}).  This is a
problem far greater than in linear theory in terms of extracting
the physics of non-linear effects, simply because it is far harder
to intuit the physics by considering a selection of different
gauges; the equations are just too complicated, and the range of
possible gauges vast. So even choosing a specific gauge at first
and second order, one has no guarantee that one has fully
understood the problem. The key problem is that for an object to
be gauge-invariant at \emph{second-order} it must vanish in the
background \emph{and} at first-order (or be constant)~\cite{BS}.
This problem has hints of a solution in the recent work of
Nakamura~\cite{nakamura}, using the methods given in~\cite{BS}, in
which a general formalism for deriving GI quantities at
second-order, assuming the existence of a method at first-order,
is given. Thus it is possible that a fully gauge-invariant
second-order theory may be presented in the near future using
Bardeen's method for constructing first-order GI quantities
together with Nakamura's approach. There has been recent progress
along these lines;~\cite{ABMR2} have given a gauge-invariant
definition of the comoving curvature at second-order for scalar
perturbations. Nevertheless, it is likely that when such a theory
is found it will be very difficult to understand exactly what
everything means, especially as the construction is not unique,
although it will likely be crucial for numerical prediction of
observable quantities such as the CMB power spectrum.

In this paper we discuss an alternative route to second-order
gauge-invariant perturbations in cosmology following the covariant
route initiated in~\cite{EB}. The covariant 1+3 approach is a good
method for perturbation theory simply because it is manifestly
covariant; many of the usual gauge-problems arise as a result of
coordinate transformations at each order. Once we have a complete
set of GI variables, then we know that they are
observable~\cite{BS}, and correspond to physical or geometrical
quantities.\footnote{It should be noted that there is a freedom
inherent in the 1+3 approach, even after a complete set of GI
variables have been specified, and this freedom lies in ones
choice of observers at the relevant perturbative order: but this
is {not} a gauge choice in perturbation theory, as it has nothing
to do with ones mapping between the background and perturbed
spacetime; it is simply a choice of observers in the perturbed
spacetime, a choice one must make in any non-vacuum spacetime. For
example, the gauge-invariant density gradient at first-order,
${\cal D}_a$, can be set to zero by a first-order Lorentz boost,
at the expense of introducing acceleration, say. Under a
zeroth-order boost, however, ${\cal D}_a$ would no longer be
gauge-invariant as it would not vanish in the background, and
would correspond to unusual observers in the background. In
metric-based approaches (where the components of the metric are
explicitly solved for), one must specify a perturbed velocity
\emph{relative to} the background, as well as having this frame
freedom. This is because the Stewart-Walker lemma and its
generalisations refer to the principle of general covariance, and
not the principle of relativity: different observers measure
different values for physical quantities such as the energy
density, the magnetic field or the electric part of the Weyl
tensor, for example. The only objects which are observer (or
tetrad) independent are the metric, the Riemann tensor (and its
trace and trace-free parts) and the Maxwell tensor.}

Using a zeroth-order background of a dust FLRW model with zero
cosmological constant, we discuss `mode-crossing' perturbations;
that is, perturbations at second order generated from an
`orthogonal mode' at first order~-- second-order  tensor modes
generated by first-order scalar modes (Section~\ref{sec1}), and
viceversa (Section~\ref{sec2}). \footnote{We ignore rotational
modes throughout. Only if they are included at first-order will
rotational modes occur which do not occur at first-order, in the
absence of acceleration.} We define a complete set of second-order
gauge-invariant quantities which fully describe the induced modes,
and convert all products of first-order quantities appearing in
the second-order equations into new second-order variables, thus
making the equations explicitly GI at second order, in a manner
very similar to~\cite{CMBD}. This simplifies the presentation of
the equations by converting them into a \emph{linear} system of
DE's, a procedure which also simplifies their solution. In the
case of first-order tensor modes, this system is infinite
dimensional reflecting the coupling of modes of each wavelength
feeding into the induced scalar modes. The standard zeroth-order
harmonic functions can be used on second-order variables to remove
the tensorial nature of the equations, and remove spatial
gradients so converting the system into odes which may then be
easily integrated. We explicitly integrate both cases under
investigation to show that scalar-mode coupling effects may play
an important role at both early and late times, and on small and
large scales.

\section{\textsc{\textmd{{first-order perturbation: the covariant formulation}}}}

The 1+3 approach relies upon the introduction of a family of
observers travelling on a four-velocity $u^a$, with which all
geometrical and physical objects and operators~-- essentially the
Riemann curvature tensor and the covariant derivative~-- are
decomposed into invariant parts; scalars along $u^a$ and scalars,
3-vectors, and projected, symmetric and trace-free (PSTF) tensors
orthogonal to $u^a$, as well as an evolution derivative along
$u^a$ and a spatial derivative orthogonal to it. The Einstein
field equations are supplemented by the Ricci identities for $u^a$
and the Bianchi identities, forming an complete set of first-order
differential equations. We refer to~\cite{EvE} for details and
references; we also follow their notation and sign
conventions\footnote{We use the standard notation whereby a dot
represents differentiation along the observers' four-velocity
$u^a$, $\dot \psi _{a\cdots b}\equiv u^c\del_c \psi _{a\cdots b}$,
and $\sdel_a$ is a derivative in the rest space of the observers;
$\sdel_c \psi _{a\cdots b}\equiv h_{c}^{~d} h_a^{~e}\cdots
h_b^{~f}\del_d \psi _{e\cdots f}$, where $h_{ab}\equiv
g_{ab}+u_au_b$ is the usual projection tensor orthogonal to $u^a$.
We use angled brackets on indices to donate the projected,
symmetric and trace-free part of a tensor. We define the
divergence of a vector $\psi_a$ as $\div\psi=\sdel^a \psi_a$, and
of a PSTF tensor $\psi_{ab}$ as $\div\psi_a=\sdel^b\psi_{ab}$; and
we define the curl of a vector as
$\curl\psi_a=\epsilon_{abc}\sdel^b\psi^c$, and of a PSTF tensor,
$\curl\psi_{ab}=\epsilon_{cd\<a}\sdel^c\psi_{b\>}^{~~d}$;
$\epsilon_{abc}$ is the observers' rest-space volume element.}.

Perturbations at first-order around a FLRW background are
relatively straightforward: define the \emph{first-order
gauge-invariant} (FOGI) variables corresponding to the spatial
fluctuations in the energy density and expansion: \ba
X_a&=&\sdel_a\mu,\\
Z_a&=&\sdel_a\theta. \ea These are FOGI simply because they vanish
in the exact FLRW background~\cite{SW,BS}. We shall not use the
more physically motivated variables ${\cal D}_a=a\sdel_a\mu/\mu$
and ${\cal Z}_a=a\sdel_a\theta$, as these tend to make the
equations more complicated to derive at second-order. Ignoring
rotational perturbations the set of FOGI equations consists of the
\emph{evolution equations} \ba
\dot X_a&=&-\:43\theta X_a-\mu Z_a,\label{ev1}\\
\dot Z_a&=&-\theta Z_a-\:12 X_a,\\
\dot \sigma_{ab}&=&-\:23\theta\sigma_{ab}-E_{ab},\label{sigdotlin}\\
\dot E_{ab}-\curl H_{ab}&=&-\:12\mu\sigma_{ab}-\theta E_{ab},\\
\dot H_{ab}+\curl E_{ab}&=&-\theta H_{ab}; \ea and the
\emph{constraints} \ba
0&=&\:13 X_a-\div E_a,\\
0&=&\:23 Z_a-\div \sigma_a,\\
0&=&\div H_a,\\
0&=&H_{ab}-\curl\sigma_{ab},\label{hlin}\\
0&=&\curl X_a,\\
0&=&\curl Z_a.\label{con6} \ea The whole system of equations is
governed \emph{entirely} by the shear, which obeys the covariant
wave-like equation: \be
\ddot\sigma_{ab}+\curl\curl\sigma_{ab}+\:53\theta\dot\sigma_{ab}
+\bra{\:49\theta^2-\:56\mu}\sigma_{ab}=0.\label{sigwavelin} \ee
Each other object is determined by solutions of this equation~--
no further integration of the equations is required after the
solution for the shear is found. Indeed it is useful to use
$\{\sigma_{ab},\dot\sigma_{ab}\}$ as the `basis vector' for the
full solution, as we shall do later. While each variable in the
problem may be shown to satisfy a wave-like equation of some sort,
these do not necessarily close; in particular, the wave equation
for $E_{ab}$ in the case of gravity wave propagation has a source
term from the shear~\cite{chal,DBE}.

Because the background is homogeneous and isotropic, each FOGI
vector may be uniquely split into a \emph{curl-free} and
\emph{divergence-free} part, usually referred to as scalar and
vector parts respectively, which we write as \be
V_a={V\S}_a+{V\V}_a~~~\mbox{where}~~~\curl
{V\S}_a=0~~~\mbox{and}~~~\div{V\V}=0. \ee Similarly, any tensor
may be invariantly split into scalar, vector and tensor parts: \be
T_{ab}={T\S}_{ab}+{T\V}_{ab}+{T\T}_{ab}~~~\mbox{where}~~~\curl
{T\S}_{ab}=0,~~~\div\div{T\V}=0,~~~\mbox{and}~~~\div{T\T}_a=0. \ee
In the equations we can  separately equate scalar, vector and
tensor parts. As we are ignoring rotation, all vector parts are
zero.

Each FOGI vector consists of only the scalar part and is
curl-free; so \be X^a={X\S}^a, \ee similarly for $Z^a$. Each
tensor consists of a scalar and a tensor part, which describe
clustering matter and gravity waves respectively. Thus the
equations can still be manipulated further: splitting the shear
into its curl free and divergence free parts gives us an
oscillatory equation for the scalar part, which tells us the
increase in shear found as matter clusters: \be
\ddot{\sigma\S}_{ab}+\:53\theta\dot{\sigma\S}_{ab}
+\bra{\:49\theta^2-\:56\mu}{\sigma\S}_{ab}=0,\label{scalsiglin}
\ee and a wave equation for the tensor part, which is the equation
governing gravity waves in an expanding dust universe: \be
\ddot{\sigma\T}_{ab}-\sdel^2{\sigma\T}_{ab}+\:53\theta\dot{\sigma\T}_{ab}
+\bra{\:19\theta^2+\:16\mu}{\sigma\T}_{ab}=0,\label{sigwavetensfo}
\ee which may be shown using the identity for a FOGI tensor: \be
\curl\curl T_{ab}=-\sdel^2T_{ab}+\:32\sdel_{\<a}\div
T_{b\>}+\bra{\mu-\:13\theta^2}T_{ab}. \ee The other two tensors
can be similarly split. Because the magnetic Weyl curvature is the
curl of the shear, the scalar part of it vanishes at this order.

We shall use Eqs.~(\ref{scalsiglin}) and~(\ref{sigwavetensfo}) to
form the source for second-order perturbations.

\section{\textsc{\textmd{{second-order perturbation}}}}

Lets consider non-linear perturbations up to second order in the
shear, ignoring rotation and fluid modes at both first and second
order, but keeping both scalar and tensor modes. The shear `wave'
equation becomes \be
\ddot\sigma_{ab}+\curl\curl\sigma_{ab}+\:53\theta\dot\sigma_{ab}
+\bra{\:49\theta^2-\:56\mu}\sigma_{ab}=\sigma_{c\<a}\dot\sigma_{b\>}^{~~c}
+\theta\sigma_{c\<a}\sigma_{b\>}^{~~c},\label{sigwavenl} \ee where
the `curl-curl' operator now has the quadratic contributions when
converting to the Laplacian: \be \curl\curl
T_{ab}=-\sdel^2T_{ab}+\:32\sdel_{\<a}\div
T_{b\>}+\bra{\mu-\:13\theta^2}T_{ab}+3E_{c\<a}T_{b\>}^{~~c}
-\theta\sigma_{c\<a}T_{b\>}^{~~c}.\label{curlcurlcom1} \ee In
deriving Eq.~(\ref{sigwavenl}) we have  substituted for the Weyl
curvature in terms of the shear, given by Eqs.~(\ref{sigdotlin})
and~(\ref{hlin}). We have neglected terms ${\cal O}(\sigma^3)$
(and derivatives of), which is a consistent covariant
linearisation procedure to second order. There are some important
things to note about Eq.~(\ref{sigwavenl}):
\begin{enumerate}
\item It is not actually second order in the usual perturbative sense.
The shear appearing in the quadratic shear terms on the rhs should
not be the same as that on the left; on the left we have a
\emph{mixture} of first and second-order quantities, but on the
right, the terms are made up of products of first-order variables.
Integrating Eq.~(\ref{sigwavenl}) without explicitly taking into
account this distinction will give a different solution than
explicitly making it so [by using solutions of
Eq.~(\ref{sigwavelin}) for the terms on the right].
\item It is not gauge-invariant.
The mixture of first and second-order quantities on the left is
really the core of the gauge problem: the `first-order bits' don't
cancel out upon application of Eq.~(\ref{sigwavelin}) because the
derivative operators in each of the two equations are not the
same. In the second-order equation there are derivative operators
of order one hanging around; but solution of the linear problem
does not tell us what these are. Hence, the equation can't be
integrated.

\end{enumerate}

We shall now convert Eq.~(\ref{sigwavenl}) into a gauge-invariant
equation at second order in the two cases we have mentioned.

\subsection{\textmd{\MakeUppercase{\sl %second-order scalar modes:
gravity waves from
density fluctuations}}}\label{sec1}

Consider the case where we excite only the scalar modes at linear
order, which describe density perturbations. These modes are
covariantly characterised at linear order by the GI condition \be
\curl\sigma_{ab}=0. \ee Hence, we may define the  variable \be
\Sigma_{ab}=\curl{\sigma}_{ab} \ee which is \emph{second-order and
gauge-invariant up to and including second-order} (SOGI), as it
vanishes at all lower orders~\cite{BS}; this variable forms the
core of the analysis of gravity waves generated by density
fluctuations. In fact, it is just the magnetic part of the Weyl
tensor if rotational modes are ignored (as may be seen from the
$\curl\sigma_{ab}$ constraint); nevertheless, it is a distinction
worth keeping, as this would not usually be the case. Taking the
curl of Eq.~(\ref{sigwavenl}), or taking the time derivative of
the magnetic Weyl evolution equation results in the equation for
$\Sigma_{ab}$: \be
\ddot\Sigma_{ab}+\curl\curl\Sigma_{ab}+\:73\theta\dot\Sigma_{ab}
+\bra{\theta^2-\mu}\Sigma_{ab}={\cal S}_{ab}\label{Sigmanl} \ee
where the source term is given by \ba {\cal
S}_{ab}&=&\epsilon_{cd\<a}\bigl\{-\:52\dot\sigma_{b\>}^{~~d}\div\sigma^c
-\:52\sigma_{b\>}^{~~d}\div\dot\sigma^c
-\:93\theta\sigma_{b\>}^{~~d}\div\sigma^c\nonumber\\&&
-\dot\sigma^{ec}\sdel_{|e|}\sigma_{b\>}^{~~d}
-2\sigma^{ec}\sdel_{|e|}\dot\sigma_{b\>}^{~~d}
+\:12\sigma^{de}\sdel^c\dot\sigma_{b\>e}
+\:12\dot\sigma^{de}\sdel^c\sigma_{b\>e}\nonumber\\&&
+\:12\sigma^{~~e}_{b\>}\sdel^c\dot\sigma_{e}^{~d}
+\:12\dot\sigma^{~~e}_{b\>}\sdel^c\sigma_{e}^{~d}
+\theta\sdel^c\bra{\sigma_{b\>e}\sigma^{de}} \bigr\}. \ea This
source term is considerably more untidy than that of the
gauge-dependent equation for the shear, Eq.~(\ref{sigwavenl})
because of the extra terms which arise from successive
applications of the commutation relation \be \bra{\curl
T_{ab}}^{^\mathbf{_\cdot}}-\curl\dot T_{ab}=-\:13\theta\curl
T_{ab}
-\epsilon_{cd\<a}\sigma^{ec}\sdel_{|e|}T_{b\>}^{~~d}\label{curldotcomm1},
\ee which holds for this case.

The second-order equation~(\ref{Sigmanl}) is not yet properly SOGI
because the rhs contains terms which are FOGI, although the rhs as
a whole is second-order, and ${\cal S}_{ab}$ is a SOGI tensor. To
make the source explicitly composed of  SOGI quantities, we define
the SOGI variables: \ba
\psi_{1ab}&=&\epsilon_{cd\<a}\bra{\div\sigma^c}\sigma_{b\>}^{~~d},\\
\psi_{2ab}&=&\epsilon_{cd\<a}\bra{\div\sigma^c}\dot\sigma_{b\>}^{~~d},\\
\psi_{3ab}&=&\epsilon_{cd\<a}\bra{\div\dot\sigma^c}\sigma_{b\>}^{~~d},\\
\psi_{4ab}&=&\epsilon_{cd\<a}\bra{\div\dot\sigma^c}\dot\sigma_{b\>}^{~~d};.
\ea It is straightforward to show that, to second order, these
variables satisfy the closed system of evolution equations \be
\dot\b\psi_{ab}=\b{\Psi\psi}_{ab},\label{psiev} \ee where
$\b\psi_{ab}=\bra{\psi_{1ab},\psi_{2ab},\psi_{3ab},\psi_{4ab}}$,
and the \emph{scalar-scalar coupling matrix} is given by \be
\b\Psi=\bra{\begin{array}{cccc}
  -\:13\theta    &  1   &  1 &0  \\
   -\:49\theta^2+\:56\mu~~   &  -2\theta   &  0  & 1\\
   -\:49\theta^2+\:56\mu~~       & 0    & -2\theta  &1\\
    0 &-\:49\theta^2+\:56\mu ~~ & -\:49\theta^2+\:56\mu ~~  &-\:{11}{3}\theta
\end{array}},
\ee which is derived using Eq.~(\ref{scalsiglin}). Similarly, we
may define the variables \ba
\phi_{1ab}&=&\epsilon_{cd\<a}(\dis\sigma^c_{~b\>e})\sigma^{de},\\
\phi_{2ab}&=&\epsilon_{cd\<a}(\dis\sigma^c_{~b\>e})\dot\sigma^{de},\\
\phi_{3ab}&=&\epsilon_{cd\<a}(\dis\dot\sigma^c_{~b\>e})\sigma^{de},\\
\phi_{4ab}&=&\epsilon_{cd\<a}(\dis\dot\sigma^c_{~b\>e})\dot\sigma^{de},
\ea where we have defined the \emph{distortion} of $\sigma_{ab}$
as \be \dis\sigma_{cab}=\sdel_{\<c}\sigma_{ab\>}, \ee which is the
remaining invariantly defined part of the spatial derivative of a
PSTF tensor which is not part of the divergence or
curl~\cite{MES}. \iffalse \ba
\phi_{1ab}&=&\epsilon_{cd\<a}\sigma^{ec}\sdel_{|e|}\sigma_{b\>}^{~~d},\\
&\vdots&\nonumber\\
\chi_{1ab}&=&\epsilon_{cd\<a}\sigma^{de}\sdel^c\sigma_{b\>e},\\
&\vdots&\nonumber\\
\eta_{1ab}&=&\epsilon_{cd\<a}\sigma^{~~e}_{b\>}\sdel^c\sigma_{e}^{~d},\\
&\vdots&\nonumber \ea\fi It is clear that these variables,
$\b\phi_{ab}$,
% $\b\chi$ and $\b\eta$,
obey the same evolution equation as $\b\psi_{ab}$, given by
Eq.~(\ref{psiev})\iffalse\footnote{The reason that four different
types of variables, of the form $\sigma\times\sdel\sigma$, are
required for ${\cal S}_{ab}$, can be seen as follows: there are
two independent spatial derivatives of the shear at first order:
the divergence, $\div\sigma_{ab}$, and the volume preserving
distortion, $\sdel_{\<a}\sigma_{bc\>}$ (see e.g.,~\cite{MES} for a
discussion of this derivative~-- we have not performed an
irreducible split here), and two different `cross products' which
may be formed with $\sigma_{ab}$, giving $2\times2=4$
variables.}\fi. By performing a complete 1+3 split of the
derivatives of the shear in the source of Eq.~(\ref{Sigmanl}) to
write them in terms of the divergence and distortion only, we may
write our source term in terms of these SOGI variables: \ba
%{\cal S}_{ab}&=&-\:93\theta\psi_{1ab}-\:53\psi_{2ab}-\:52\psi_{3ab}
%-\phi_{2ab}-2\phi_{3ab}\nonumber\\&&
%+2\theta\chi_{1ab}+\:12\chi_{2ab}+\:12\chi_{3ab}
%+2\theta\eta_{1ab}+\:12\eta_{2ab}+\:12\eta_{3ab}.\\
 {\cal S}_{ab}&=&-\:{22}{5}\theta\psi_{1ab}-\:{27}{20}\psi_{2ab}
 +\:{3}{20}\psi_{3ab} +\theta\phi_{1ab} +\:{3}{2}\phi_{2ab}
 +\:{5}{2}\phi_{3ab}.
\ea We now have a complete closed linear system of differential
equations which may be expanded in the usual background harmonics,
and integrated straightforwardly (especially numerically) once
initial data is prescribed. By taking various spatial gradients of
these eight variables it can be shown that there are some
(differential) constraints which must be satisfied amongst these
variables, which serve to constrain initial data, but we shall not
pursue this further here (see~\cite{SMEL} for details of where
these come from). Note that we have managed to deal with all the
complicated scalar-scalar coupling without recourse to harmonics.

The mode-mode coupling variables, $\b\psi_{ab},\b\phi_{ab}$ do
several things in aiding the solution to Eq~(\ref{sigwavenl}) at
second-order:  they transform Eq.~(\ref{Sigmanl}) into a genuinely
second-order equation in the perturbative sense because in
calculating their evolution equations we neglected terms ${\cal
O}(\sigma^3)$, and made the final distinction between the
variables on the lhs and those on the right; they turn
Eq.~(\ref{Sigmanl}) into an explicitly SOGI equation consisting
only of SOGI variables; and finally they turn the system of
equations which must be solved into a system of \emph{linear}
differential equations, which are much easier to integrate
numerically, and analyse by other techniques, than systems which
are not linear.

\subsubsection{\textmd{\textsc{the induced tensor modes}}}

The mode coupling which provides the source for the second order
perturbation, although being products of scalar mode variables,
induces tensor modes at second order (and, in  more general
situations, rotational modes), seen by the fact that $\curl {\cal
S}_{ab}\neq0$. This mode-mixing is perhaps one of the most
interesting parts of non-linear perturbation theory; as we  show
here, for example, gravity waves may be generated~-- or reduced~--
purely from the effects of linear clustering of matter. The
second-order scalar modes which $\Sigma_{ab}$ tells us about,
however, are not genuine second order effects: for scalar modes,
$\div\Sigma_a\neq0$, and is in fact just a combination of
quadratic first-order variables, as may be seen through the $\div
H_a$ constraint. It is interesting to note that \be
\div\Sigma_a=\epsilon_{abc}\sigma^{bd}\dot\sigma^c_{~d}=\sum_{k,k'}\sigma_{(k)}\dot\sigma_{(k')}
 \underbrace{\epsilon_{abc}Q_{(k)}^{bd}{Q_{(k')}}^c_{~d}}_{=0~\mbox{when}~k= k'},
\ee using standard harmonics~\cite{chal,EvE}, so coupling between
modes of differing wavelength do not give rise to tensor modes at
second order, for which $\div\Sigma_a=0=\div H_a$. This is a
reflection of the integrability conditions discussed
in~\cite{SMEL}, which suggests that $\div H_a=0$ may not be as
restrictive in perturbative models as it is in the fully
non-linear case.

The second order tensor modes, on the other hand, are covariantly
characterised by $\div\Sigma_a=0$, for which we have the genuine
wave equation:
\be
%\ddot\Sigma_{ab}-\sdel^2\Sigma_{ab}+\:73\theta\dot\Sigma_{ab}
%+\bra{-\:29\theta^2+\:53\mu}\Sigma_{ab}={\cal S}_{ab}\label{Sigmawavetensor},
{{{\ddot\Sigma}{\T}}}_{ab}-\sdel^2{\Sigma\T}_{ab}+\:73\theta{\dot\Sigma{\T}}_{ab}
+{\:23\theta^2}{\Sigma\T}_{ab}={\cal
S}{\T}_{ab}\label{Sigmawavetensor},
\ee
upon using Eq.~(\ref{curlcurlcom1}) in Eq.~(\ref{Sigmanl}).
Harmonic functions, defined as usual in the background (any
changes would be ${\cal O}(3)$), may now be introduced, and the
equations integrated in time; we discuss this in the case of a
flat background below.

Now, gravity waves are waves in the Weyl curvature which
\emph{induce} waves in the shear (non-local waves which travel at
the speed of light and thus can't be removed by a change of
observers), so in order to fully understand gravitational
radiation induced by density fluctuations, we must relate
$\Sigma_{ab}$ to the Weyl curvature. This simply involves creating
SOGI variables for the Weyl curvature. The magnetic part is easy,
as we have $H_{ab}=\Sigma_{ab}$ in the absence of rotation, while
for the electric part of the Weyl tensor we may define
\be
\mathscr E_{ab}=\curl E_{ab}=-\dot\Sigma_{ab}-\theta\Sigma_{ab}
%-\psi_{1ab}-\phi_{1ab}-\chi_{1ab}-\eta_{1ab},
+\:65\psi_{1ab},
\ee
which completes the solution for the induced tensor modes.

In the case of a flat background, the equations are particularly
easy to integrate (see~\cite{chal} for details on solving the
homogeneous part of Eq.~(\ref{Sigmawavetensor}) in a non-flat
background): we have the scale factor $a=a_0t^{2/3}$ (we set
$a_0=1$), implying $\theta=2/t$ and $\mu=\:13\theta^2$;
integrating Eq.~(\ref{psiev}) for $\b\psi_{ab}$ and $\b\phi_{ab}$,
we find that the source has three components:
\be
{\cal S}_{ab}=\alpha_{ab}
t^{-7/3}+\beta_{ab}t^{-4}+\gamma_{ab}t^{-17/3},\label{sourcetens}
\ee
where the last term is the largest at early times (which has the
time dependence of $a^{-1}\sigma\dot\sigma$). The tensors
$\alpha_{ab},~\beta_{ab}$ and $\gamma_{ab}$ are constant in time
and are a combination of the initial conditions of $\b\psi_{ab}$
and $\b\phi_{ab}$; the details of this need not concern us here.
Using standard tensor harmonics~\cite{BDE,chal} removes the
tensorial nature and spatial dependence of the wave
equation~(\ref{Sigmawavetensor}), and replaces the covariant
Laplacian with a harmonic index, $\sdel^2\mapsto-k^2/a^2$, the
full solution being a sum over harmonic modes. The analytic
solution to Eq.~(\ref{Sigmawavetensor}) may be easily obtained:
\be
\Sigma^{(k)}=\Sigma_1^{(k)}+\Sigma_2^{(k)}+\alpha^{(k)}k^{-2}t^{-1}
+\beta^{(k)}k^{-2}t^{-8/3}+\gamma^{(k)}\times{\mbox{mess}}
\ee
where the two solutions to the homogeneous part are
\ba
\Sigma_1^{(k)}&=&S_1^{(k)} t^{-8/3}\bras{\bra{3x^2-1}\cos 3x-3x\sin3x}\\
\Sigma_2^{(k)}&=&S_2^{(k)} t^{-8/3}\bras{\bra{3x^2-1}\sin
3x+3x\cos3x}
\ea
where $x=kt^{1/3}$ and $S_1^{(k)},S_2^{(k)}$ are constants. The
`mess' from the $\gamma_{ab}$ term dominates the solution at early
times and on super-Hubble scales, when $t\ll 1$ and $x\ll 1$
respectively: \be \Sigma^{(k)}\sim \:38\gamma^{(k)}
t^{-11/3}\bras{1-\:34 x}+{\cal O}({t^{-8/3}}), \ee with the next
term arising from the homogeneous solution. Meanwhile, at late
times we have a new power law behaviour dominating from the scalar
mode interaction \be \Sigma^{(k)} \sim  k^{-2}\alpha^{(k)} t^{-1}
+3k^2t^{-2}\brac{S_2^{(k)}\sin3x-\bra{S_1+\:{81}{256}k^3\pi
\gamma^{(k)}}\cos3x}
+{\cal O}(t^{-7/3}),% +\beta^{(k)} t^{-8/3} }.
\ee while we see that on small scales ($k\gg1$) at late times the
oscillatory behaviour is also dominated by the scalar mode
interaction. Thus, in many regions of interest the terms arising
from the scalar mode coupling dominate the solution.

\begin{figure}[ht!]
\center\includegraphics[width=0.8\textwidth]{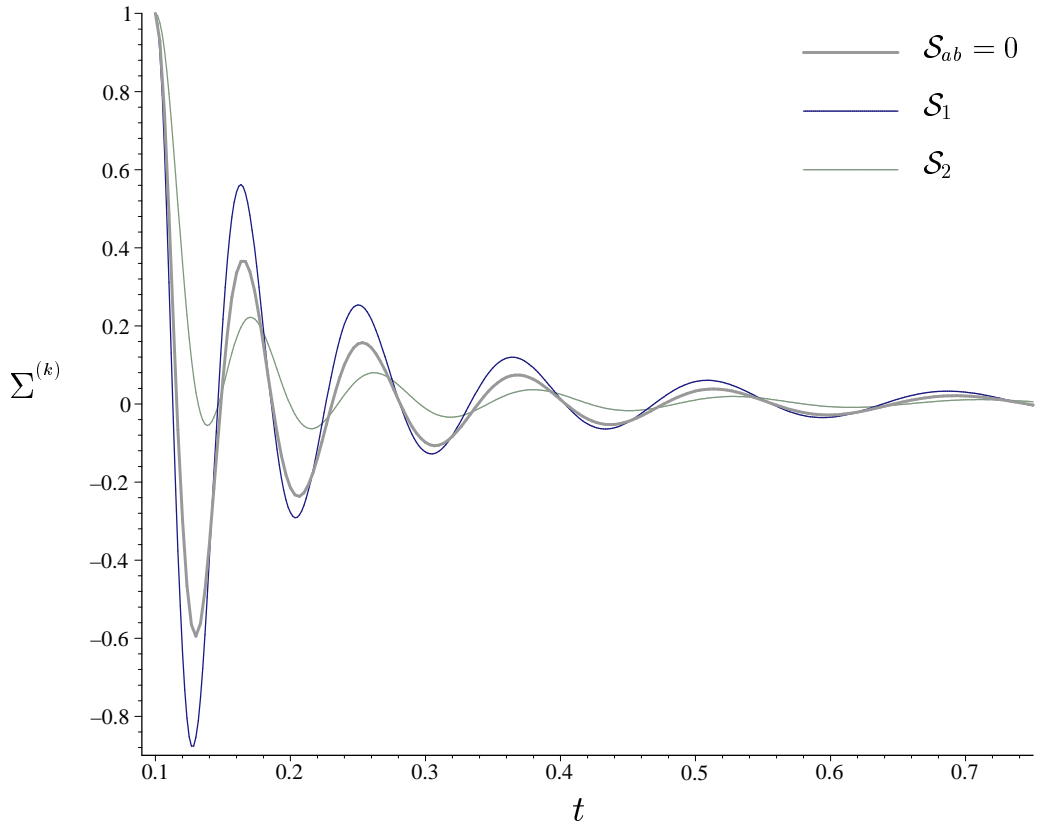}
\caption{\small The induced gravity waves from non-linear density
fluctuations for a flat background. Here we have shown the
harmonic component of $\Sigma_{ab}$, $\Sigma^{(k)}$, corresponding
to $k/a_0=25$ in three situations, all of which have initial
conditions such that $\Sigma^{(k)}=1$ and $\dot\Sigma^{(k)}=0$ at
$t=0.1$ (in arbitrary units). The thick grey curve  represents the
oscillatory decay of gravity waves with no source present, as
predicted from linear theory. To contrast with that we have shown
two different cases: curve ${\cal S}_1$ has parameters from the
source term $\alpha^{(k)}=0,~\beta^{(k)}=0.8,~\gamma^{(k)}=-.03$,
which shows considerable amplification of the gravity waves over
the linear case, while curve ${\cal S}_2$ has
$\alpha^{(k)}=1,~\beta^{(k)}=0,~\gamma^{(k)}=.02$, which shows
significant damping.
\label{nonlinearGW-from-dust-FLRW-scalar-perts_LAB.ps}}
\end{figure}

In Fig.~\ref{nonlinearGW-from-dust-FLRW-scalar-perts_LAB.ps} we
show some solution curves for $\Sigma^{(k)}$ for a flat model both
with and without the non-linear source. This shows that the effect
of including the scalar-scalar interaction in the generation of
gravity waves may increase or decrease the amplitude and power of
the induced tensor perturbations for a given wavelength, possibly
quite significantly. This may have implications for the spectrum
of tensor modes in the cosmic microwave background (see
e.g.,~\cite{chal,MM97} for details of calculating the CMB power
spectrum from the solutions given here).
%Indeed, recent results
%from the WMAP measurements~\cite{WMAP} of the CMB power spectrum
%indicate that gravitational waves are considerably weaker than has
%been predicted from many inflationary theories; this mode-mode
%coupling we have under investigation may provide part of the
%explanation for this.

\subsection{\textmd{\MakeUppercase{\sl %second-order tensor modes:
density fluctuations from gravity waves }}}\label{sec2}

We shall now look at the reverse situation from the previous
section, and consider the case where only tensor perturbations are
excited at first order. To second order we shall still neglect
rotation and fluid modes for simplicity of exposition, and study
the induced density fluctuations, characterised by the scalar
modes.

Pure tensor modes are covariantly characterised by the FOGI
condition \be \div\sigma_a=0 \ee from which it follows that the
divergence of the two Weyl curvature tensors must vanish also.
Thus, at second order, we use the SOGI variable
\be
\Delta_a=\div\sigma_a
\ee
as our key variable. From the $\div H_a$-constraint and the
non-linear commutation relation
\be
\div\curl T_a-\:12\curl\div T_a=\epsilon_{abc}E^{bd}T^c_{~d}
-\:13\theta\epsilon_{abc}\sigma^{bd}T^c_{~d},\label{divcurlcomm}
\ee
we find that $\curl\Delta_a=0$ because we're ignoring rotation~--
so $\Delta_a$ contains only scalar modes. Our wave-like equation
for $\sigma_{ab}$, Eq.~(\ref{Sigmanl}), becomes, upon taking its
divergence,
\be
\ddot\Delta_a+\:73\theta\dot\Delta_a+\bra{\theta^2-\mu}\Delta_a={\cal
T}_{a} \label{Deltawave}
\ee
where the source arising from tensor-tensor coupling is,
neglecting terms ${\cal O}(\sigma^3)$ and ${\cal
O}(\sigma\times\Delta)$,
\be
{\cal T}_{a}=-\:23\theta\sdel_a\bra{\sigma^{bc}\sigma_{bc}}
%\:73\theta\sigma^{bc}\sdel_b\sigma_{ac}
-\:43\sdel_a\bra{\sigma_{bc}\dot\sigma^{bc}},
\ee
which is remarkably simple, despite several applications of the
commutation relation Eq.~(\ref{divcurlcomm}) and \be \bra{\div
T_{a}}^{^\mathbf{_\cdot}}-\div\dot T_{a}=-\:13\theta\div T_{a}
-\sigma^{bc}\sdel_bT_{ac}+
\epsilon_{abc}H^{bd}T^c_{~d}\label{divdotcomm1}. \ee We have again
substituted for the Weyl curvature in terms of the shear.

As in the previous case, we may make the source for
Eq.~(\ref{Deltawave}) explicitly SOGI by introducing a set of SOGI
variables; variables which are constructed from quadratic FOGI
variables. However, the presence of the Laplacian in the wave
equation for the tensor modes at first-order means that we need an
infinite set of these variables; a mode of one particular
wavelength at first-order couples with all higher wavelength modes
at second order. One can approach this in two ways: explicitly
split the first-order solution into a sum over tensor harmonics;
when inserted into the source term ${\cal T}_a$ we have an
infinite sum (over $k$ and $k'$) over terms of the form
$\sigma^{(k)}\sigma^{(k')}Q\T^{(k)}Q\T^{(k')}$, each of which may
be expanded as an infinite sum over tensor harmonics. Or, one can
covariantly define an infinite set of SOGI quadratic variables,
without expanding the first-order solution in harmonics. Both
methods are more or less the same when the tensor harmonics form a
complete set of basis functions over which any (smooth) solution
of the first-order wave equation~(\ref{sigwavetensfo}) may be
expanded. However, there are certain situations in which the
first-order solution may not be completely given as a sum over
harmonic modes~-- black hole perturbation theory being perhaps the
most widely known case~\cite{nollert}\footnote{Expanding the key
wave equation for first-order gravity waves around a black hole in
temporal harmonics does not sum to the complete solution, because
of purely outgoing boundary conditions in  that problem.}.
Therefore we will explore the second route, although the first may
be more suitable for many applications.

We define the variables \ba
\psi_{1a}^{(m,n)}&=&a^{2n+2m+1}\sdel_a\bras{\sdel^{(2m)}\sigma_{bc}
\sdel^{(2n)}\sigma^{bc}},\nonumber\\
\psi_{2a}^{(m,n)}&=&a^{2n+2m+1}\sdel_a\bras{\sdel^{(2m)}\dot\sigma_{bc}
\sdel^{(2n)}\sigma^{bc}},\nonumber\\
\psi_{3a}^{(m,n)}&=&a^{2n+2m+1}\sdel_a\bras{\sdel^{(2m)}\sigma_{bc}
\sdel^{(2n)}\dot\sigma^{bc}},\nonumber\\
\psi_{4a}^{(m,n)}&=&a^{2n+2m+1}\sdel_a\bras{\sdel^{(2m)}\dot\sigma_{bc}
\sdel^{(2n)}\dot\sigma^{bc}}, \ea where the indices $m,n$ are
positive integers (including zero), and $\sdel^{(2n)}$ represents
the Laplacian operating $n$-times. Note that $\psi_1$ and $\psi_4$
are symmetric on the $m$ and $n$ indices, while
$\psi_{2a}^{(m,n)}=\psi_{3a}^{(n,m)}$~-- these forming a set of
constraints. The variables evolve, to second order, as \be
\dot\b\psi_a^{(m,n)}=\b\Psi\b\psi_a^{(m,n)}+\b\lambda\b\psi_a^{(m+1,n)}
+\b\chi\b\psi_a^{(m,n+1)}\label{skljdcn} \ee where the
\emph{tensor-tensor coupling matrix} is given by \be
\b\Psi=\bra{\begin{array}{cccc}
  0    &  1   &  1 &0  \\
   -\:19\theta^2-\:16\mu~~   &  -\:53\theta   &  0  & 1\\
   -\:19\theta^2-\:16\mu~~       & 0    & -\:53\theta  &1\\
    0 &-\:19\theta^2-\:16\mu ~~ & -\:19\theta^2-\:16\mu ~~  &-\:{10}{3}\theta
\end{array}},
\ee while the \emph{cascading matrices} are \be
\b\lambda=\bra{\begin{array}{cccc}
  \cdot & \cdot & \cdot & \cdot \\
  a^{-2} & \cdot & \cdot & \cdot \\
  \cdot & \cdot & \cdot & \cdot \\
  \cdot & \cdot & a^{-2} & \cdot
\end{array}},~~~~~~~~~
\b\chi=\bra{\begin{array}{cccc}
  \cdot & \cdot & \cdot & ~\cdot \\
  \cdot & \cdot & \cdot & ~\cdot \\
  a^{-2} & \cdot & \cdot & ~\cdot \\
  \cdot & a^{-2} & \cdot & ~\cdot
\end{array}},
\ee which couple modes of differing wavelength and feed them into
the source for the scalar modes. We also require the set of
variables \ba
\phi_{1a}^{(m,n)}&=&a^{2n+2m+1}\sdel^b\bras{\sdel^{(2m)}\sigma_{c\<
a}
\sdel^{(2n)}\sigma_{b\>}^{~~c}},\nonumber\\
\phi_{2a}^{(m,n)}&=&a^{2n+2m+1}\sdel^b\bras{\sdel^{(2m)}\dot\sigma_{c\<
a}
\sdel^{(2n)}\sigma_{b\>}^{~~c}},\nonumber\\
\phi_{3a}^{(m,n)}&=&a^{2n+2m+1}\sdel^b\bras{\sdel^{(2m)}\sigma_{c\<
a}
\sdel^{(2n)}\dot\sigma_{b\>}^{~~c}},\nonumber\\
\phi_{4a}^{(m,n)}&=&a^{2n+2m+1}\sdel^b\bras{\sdel^{(2m)}\dot\sigma_{c\<
a} \sdel^{(2n)}\dot\sigma_{b\>}^{~~c}}, \ea where is is clear that
these variables $\b\phi_a^{(m,n)}$ obey exactly the same evolution
and constraint equations as $\b\psi_a^{(m,n)}$.

Our source now becomes explicitly SOGI, so that the full evolution
equation for induced scalar perturbations from gravity waves
becomes very simple:
\be
\ddot\Delta_a+\:73\theta\dot\Delta_a+\bra{\theta^2-\mu}\Delta_a=
%\:73a^{-1}\theta\phi_{1a}^{(0,0)} -a^{-1}\psi_{2a}^{(0,0)}.
-\:23\theta a^{-1}\psi_{1a}^{(0,0)}-\:43a^{-1}\psi_{2a}^{(0,0)}
\label{Deltawave2}
\ee
We now have a SOGI linear system of pure evolution equations,
which may be integrated in a straightforward manner (although the
infinite-dimensionality of the system may pose a few problems;
this would typically be dealt with with a short wavelength
cutoff). There will also be a system of constraint equations
between all the $\b\phi_a^{(m,n)}$ and $\b\psi_a^{(m,n)}$ which
are derivable from their definition; these would serve to
constrain initial data when integrating the full system of
equations (or, if the system of equations were analysed in tensor
harmonics, would provide a system of algebraic equations which
could then be solved, giving a smaller system of evolution
equations to integrate).

The complete solution for scalar modes is given when
Eq.~(\ref{Deltawave2}) is integrated: the spatial gradient of the
expansion is given by the div-shear constraint \be
Z^a=\:32\Delta^a \ee while the gradient of the energy density is a
little more complicated: \be
X_a=-3\dot\Delta_a-3\theta\Delta_a+2a^{-1}\phi_{1a}^{(0,0)}+a^{-1}\psi_{1a}^{(0,0)}
\ee follows from the $\div E_a$ constraint.

We may once again integrate our key equation~(\ref{Deltawave2}) in
a flat background. As an illustration, consider only super-Hubble
tensor modes, and set $m=n=0$. The source term may be found by
integrating Eq.~(\ref{skljdcn}), and has the same time dependence
as the source in the previous case, Eq.~(\ref{sourcetens}). The
solution for $\Delta_a$ is \be \Delta_a=T_{1a} ~t^{-1}+T_{2a}
~t^{-8/3}+\underbrace{\:{9}{14}\alpha_a ~t^{-1/3}-\:32\beta_a
~t^{-2}+\:38\gamma_a ~t^{-11/3}}_{\mbox{\sc{\small induced
modes}}}, \ee which shows that the tensor mode coupling dominates
at both early and late times, and, in fact, contains a growing
mode ($\Delta_a/\theta\sim \alpha a$).

At late times density perturbations are the key for structure
formation in the observable universe, and a relativistic analysis
must be used for scales approaching the Hubble length. Density
perturbations may be directly related to the growth rate of
clustering matter~\cite{EEM}. As tensor modes may alter the
behaviour of clustering matter, evidence may be found in variables
such as the bias parameter.

\section{\textsc{\textmd{{conclusions}}}}

Understanding non-linear effects in cosmology will likely play a
pivotal role in cosmology in coming years, as observations pin
down the overall properties of the universe ever more tightly.
Most of the problems preventing a systematic study of non-linear
relativistic effects are mathematical: fully non-linear
relativistic simulations are beyond the horizon in cosmology at
present, leaving perturbation theory as our main probe. But
non-linear perturbation theory is not particulary simple as gauge
problems and large equations make it quite untidy. In this paper
we have considered for the first time the 1+3~covariant approach
to gauge-invariant non-linear perturbation theory.

The main problem  the 1+3 covariant approach has at second order
is intricately related to the gauge-problem: while second-order
equations, such as Eq~(\ref{sigwavenl}), can be written down~-- by
crossing off all the third order terms~-- they can't in general be
integrated. This is because the derivative operators, through the
commutation relations, contain lots of first order stuff, which
the covariant approach does not solve for at first-order (it only
solves for physical variables, not operators, in contrast with the
metric approach). In order to integrate the second-order equations
therefore, derivative operators (dot and $\sdel$) must operate
only on variables which vanish at first and zeroth order~-- SOGI
variables, in other words. We created these SOGI variables in two
ways: by appropriate derivatives of first-order quantities which
vanish at first-order~-- exactly the same trick used to create
FOGI variables from zeroth-order scalars; and from explicit
products of FOGI quantities.

The GI approach we develop here has the advantage of producing a
relatively simple linear system of odes, which is easy to
integrate. By integrating the SOGI equations, we have found that
non-linear effects may well play an important role at both early
and late times. Indeed in both cases we examined we found that
second-order modes will dominate the perturbation spectrum at late
times, which is a consequence of the fact that second-order modes
arise partly as an integrated effect, and can't always be
neglected by assumption. In addition, tensor modes generated by
non-linear effects may play an key role on super-Hubble scales.
This analysis suggests that linear perturbation theory may not be
a sufficiently accurate tool to understand some subtle dynamical
aspects of the universe, but further investigation is required to
find out exactly what these are.

\acknowledgments

It is a pleasure to thank Richard Barrett, Marco Bruni, George
Ellis, Antony Lewis, Roy Maartens and Bob Osano for useful
discussions and comments. This project was funded by NRF (South
Africa).


\begin{thebibliography}{90}

\bibitem{GLMM} Gangui, A., Lucchin, F., Matarrese, S. and Mollerach, S.,
Astrophys. J. {\bf 430} 447 (1994)

\bibitem{ABMR2} Acquaviva, V. Bartolo, N.,. Matrarrese, S., and Riotto, A.
Nuc. Phys. B {\bf667} 119 (2003)


\bibitem{tomita} Tomita, K. Prog. Theor. Phys. {\bf37} 831 (1967)


\bibitem{MTB} Mena, F. C., Tavakol, R. and Buni, M.  Int. J. Mod. Phys. A
{\bf17} 4239 (2002)

\bibitem{MMB} Matarrese, S., Mollerach, S. and Bruni, M. Phys. Rev. D {\bf
58} 043504 (1998)

\bibitem{MW} Malik, K. A. and Wands, D. astro-ph/0307055 (2003)

\bibitem{BCLM} Bartolo, N., Corasaniti, P-S., Liddle, A. R.,
Malquarti, M. astro-ph/0311503 (2003)

\bibitem{BMR1} Bartolo, N., Matarrese, S. and Riotto, A.
astro-ph/0308088 (2003)

\bibitem{BMR2} Bartolo, N., Matarrese, S. and Riotto, A.
astro-ph/0309692 (2003)

\bibitem{MHM} Mollerach, S., Harari, D. and Matarrese, S.
astro-ph/0310711 (2003)

\bibitem{SW} Stewart, J.M. and Walker, M. Proc. R. Soc. London A {\bf 431} 49
(1974)

\bibitem{BS} Bruni, M., Matarrese, S., Mollerach, S. and Sonego, S., Class.
Quantum Grav. {\bf 14} 2585 (1997); Bruni, M. and Sonego, S.,
 Class. Quantum Grav.,  {\bf 16} L29 (1999); Sopuerta, C. F.,
 Bruni, M., and Gaultieri, L., gr-qc/0306027 (2003)

\bibitem{bardeen} Bardeen, J. Phys. Rev. D {\bf 22} 1882 (1980)

\bibitem{EB} Ellis, G.F.R. and Bruni, M. Phys Rev. D {\bf 40} 1804 (1989)

\bibitem{BDE} Bruni, M., Dunsby, P. K. S., and Ellis, G. F. R., Astrophys. J. {\bf
395}, 34 (1992)

\bibitem{CMBD} Clarkson, C. A., Marklund M., Betschart, G. and Dunsby, P. K.
S. astro-ph/0310323 (2003)

\bibitem{EvE} G. F.~R.  Ellis and H. van Elst, in M. Lachieze-Rey (ed.),
{\em Theoretical and Observational Cosmology}, NATO Science
Series, Kluwer Academic Publishers~(1998) {\em gr-qc/9812046v4}

\bibitem{nakamura} Nakamura, K.,  Prog. Theor. Phys. {\bf110} 723 (2003)
%gr-qc/0303090 (2003)

\bibitem{chal} Challinor, A., Class Quantum Grav. {\bf 17} 871 (2000)

\bibitem{MM97} Mollerach, S. and Matarrese, S., Phys. Rev. D
{\bf56} 4494 (1997)

\bibitem{DBE} Dunsby, P. K. S., Bassett, B. A. C. C. and Ellis, G. F. R.,
Class. Quantum Grav. {\bf 14} (1997)

\bibitem{MES} Maartens, R., Ellis, G. F. R., and Siklos, S. T. C., Class.
Quantum Grav. {\bf 14} 1927 (1997)

\bibitem{SMEL} Sopuerta, C. F., Maartens, R., Ellis, G. F. R. and Lesame, W.
M. gr-qc/9809085 (1999)

\bibitem{WMAP} Bennett, et. al., Astrophys. J. Suppl. {\bf 148} 1 (2003);
 Tegmark, M. et al.,
astro-ph/0310723 (2003)

\bibitem{nollert} Nollert, HP. Class. Quantum Grav. {\bf 16} R159 (1999);
Kokkotas, K.D. and Schmidt, B.G., Living Rev. Relativity, {\bf 2}
2 (1999)

\bibitem{EEM} Ellis, G. F. R., van Elst, H. and Maartens, R. Class.Quant.Grav. {\bf18} 5115 (2001)






\end{thebibliography}
\end{document}